\documentstyle[preprint,aps,prl,epsfig]{revtex}

\begin{document}
\draft
\preprint{ISKP-Preprint 12-98-1}
\title{P-wave dominance in the
\mbox{\boldmath $pd\to\, ^{3}$}He\mbox{\boldmath $\,\pi^+\pi^-$}
reaction near threshold measured with the MOMO facility at COSY}
\author{F.~Belleman,$^1$ A.~Berg,$^1$ J.~Bisplinghoff,$^1$ 
G.~Bohlscheid,$^1$ J.~Ernst,$^1$ C.~Henrich,$^1$
F.~Hinterberger,$^1$ R.~Ibald,$^1$ R.~Jahn,$^1$
L.~Jarczyk,$^2$ R.~Joosten,$^1$ A.~Kozela,$^3$ 
H.~Machner,$^4$ A.~Magiera,$^2$ R.~Maschuw,$^1$
T.~Mayer-Kuckuk,$^1$ G.~Mertler,$^1$ J.~Munkel,$^1$
P.~von~Neumann-Cosel,$^5$ D.~Rosendaal,$^1$ P.~von Rossen,$^4$
H.~Schnitker,$^1$ K.~Scho,$^1$ J.~Smyrski,$^2$
A.~Strzalkowski,$^2$ R.~T\"olle,$^4$ and C.~Wilkin $^6$
  }

\address{
$^1$ Institut f\"ur Strahlen- und Kernphysik, Universit\"at Bonn \\
$^2$ Institute of Physics, Jagellonian University, Cracow, Poland\\
$^3$ Institute of Nuclear Physics, Cracow, Poland\\
$^4$ Institut f\"ur Kernphysik, Forschungszentrum J\"ulich\\
$^5$ Institut f\"ur Kernphysik, Technische Universit\"at Darmstadt\\
$^6$ Department of Physics and Astronomy, University College London,
London WC1E 6BT, UK
}

\date{\today}

\maketitle

\begin{abstract}
The cross section for the  $pd\to\,^{3}\mbox{\rm He}\,\pi^+\pi^-$
reaction has been measured at the MOMO facility in a kinematically complete
experiment at a c.m.\ excess energy of $Q=70$~MeV. The energy and angular
distributions show that the reaction is dominated by $p$-wave $\pi^+\pi^-$
pairs. This is in complete contrast to the results of inclusive measurements
at somewhat higher beam energies which show a strong $s$-wave ABC enhancement
at low $\pi\pi$ masses. There are however indications of $p$-wave pion pairs
from other experiments at low $Q$.\\
\end{abstract}


\narrowtext
The measurements of inclusive meson production in the  
$pd\to\, ^{3}\mbox{\rm He}\,X^0$ reaction~\cite{ABC} are
surprising, in that they show a strong enhancement for a
missing mass of around 310~MeV with a width of only 50~MeV. The fact
that no similar effect is seen for the $^3\mbox{\rm H}\,X^+$ final
state suggests that this anomaly must be associated with the isospin-zero
s-wave $\pi\pi$ system. On the other hand, since the observed mass and
width of the peak both vary with beam energy $T_p$~\cite{Banaigs}, and
it is known that the corresponding isoscalar $\pi\pi$ scattering length
is small~\cite{Grayer}, it has generally been assumed that the anomaly
must be kinematic in origin, possibly being associated with the
production of two $\Delta$-isobars in the reaction~\cite{Risser}.

These inclusive measurements were carried out for
$T_p\geq 745$~MeV~\cite{ABC,Banaigs}, corresponding to c.m.\ kinetic
energies in the final state $Q\geq 190$~MeV. To investigate such pion
production in greater detail, with the objective of deducing also
angular distributions, we have carried out an exclusive
measurement of the $pd\to\, ^{3}\mbox{\rm He}\,\pi^+\pi^-$ reaction
closer to threshold at $Q=72$~MeV ($T_p=546$~MeV).

The MOMO facility was installed at the external proton beam of the COSY
proton synchrotron of the Forschungszentrum J\"ulich~\cite{Maier}. The
setup consists of a high granularity meson vertex detector near the
target, with the high resolution 3Q2DQ magnetic spectrometer Big
Karl~\cite{Karl} being placed in the forward direction. The acceptance of
this spectrometer is $\pm 25$~mrad in the horizontal and $\pm 100$~mrad
in the vertical directions. The charged particle tracks were measured in
the focal plane by two stacks of multiwire drift chambers (MWDC) which
yield position information in both the horizontal and the vertical
directions. The $^{3}\mbox{\rm He}$'s were unambiguously identified by
their time-of-flight and energy loss, as measured with two scintillator
hodoscopes behind the MWDC's, separated 2m from each other. By itself
this gave a missing mass resolution in the 
$pd\to\, ^{3}\mbox{\rm He}\,X$ reaction of typically 1~MeV/c$^2$. In
order to make a kinematically complete identification of the events,
the directions of the pions were measured in the MOMO vertex
detector~\cite{thesis}. This consists of 672 scintillating fibers, of
circular profile with diameters of 2.5~mm, arranged in three planes
tilted at 60$^o$ with respect to each other. The fibers are individually
read out by 16-fold photomultipliers. The detector was placed
perpendicular to the beam direction at 20~cm after the target,
subtending an opening angle of $\pm 45^0$. A 4~cm diameter central hole
allowed the $^{3}\mbox{\rm He}$'s and the undeflected proton beam to pass.
A liquid deuterium target, 4~cm thick and 6~mm in diameter with 1~$\mu$m
mylar windows~\cite{target}, was placed in the MOMO vacuum chamber. A
phosphorus screen, which could be lowered directly behind the target,
showed the beam spot to be about 1~mm in diameter. A typical beam intensity
of $10^9$ particles per second was used in the experiment.

An event with two charged particles in the vertex detector and a $^3$He
in Big Karl was considered to be a candidate for the 
$pd\to\, ^{3}\mbox{\rm He}\,\pi^+\pi^-$ reaction and its identification
and complete reconstruction involved a 2C kinematic fit. Good events
must be coplanar with respect to the total meson momentum axis, which is
defined by the beam and the $^{3}\mbox{\rm He}$ momenta. The distribution
in coplanarity of the coincident hits, shown in fig.~1, demonstrates that
any background due to four-body reactions or random coincidences is at
most a few per cent.

About 15,000 fully reconstructed $pd\to\,^{3}\mbox{\rm He}\,\pi^+\pi^-$
events were obtained at a beam energy of 546~MeV. Although the overall
acceptance of the MOMO apparatus is only about 2\% of 4 $\pi$, it is well
distributed over the complete phase space, with the exception of the maximum
excitation energy of the pion pair, where at least one of the pions
escapes the vertex detector. The measurements were performed with
three settings of the Big Karl magnet and the consistency of the results
obtained in the overlapping regions shows that the acceptance of the
spectrometer is well understood.

The differential cross sections obtained are displayed in fig.~2 in
terms of four of the possible kinematic variables. The absolute 
normalization was determined using various scattering monitors calibrated 
to about $\pm 7$\%, and this is by far the largest contribution to the
systematic error. The only accessible variable in single-arm
experiments~\cite{ABC,Banaigs} is the pion-pion excitation energy
$T_{\pi\pi}= m_{\pi\pi}-2m_{\pi}$, where $m_{\pi\pi}$ is the two-pion
effective mass, and the distribution in this is shown in fig.~2a. In marked
contrast to the original ABC experiments~\cite{ABC}, which showed an
enhancement over phase space in the region of $T_{\pi\pi}\approx 30$~MeV,
our data are pushed rather to maximum values of excitation energy. This
strong energy dependence suggests that higher partial waves are
contributing. On the other hand, the distribution in the $\pi\,^3$He
relative energy, shown in fig.~2b, is fairly consistent with phase space.
This attests that the anomalous behaviour in $T_{\pi\pi}$ is
not an artefact of the MOMO acceptance. Pure $s$-waves would lead to no
dependence upon $\theta_{\pi}$, the angle between the proton and one of
the pions in the overall c.m.~system. The significant anisotropy shown
in fig.~2c is therefore direct confirmation of the importance of higher
partial waves. It should be noted that events with pions in the backward
hemisphere would not in general be detected in the MOMO apparatus.
A particularly interesting angular variable for the subsequent discussion
is that between the relative $\pi\pi$ momentum and the beam
direction in the overall c.m.\ system, and the corresponding distribution
is shown in fig.~2d. Since no distinction is made in MOMO between the
$\pi^+$ and $\pi^-$, such a distribution must be symmetric about $90^o$,
and it is striking that over most of the range the cross section
is linear in $\sin^2\theta_{\pi\pi p}$.

It is seen that the $\pi\pi$ excitation energy distribution of fig.~2a
is broadly compatible with phase space multiplied by $T_{\pi\pi}$. 
This, together with the linearity shown in fig.~2d, indicates that the
$\pi\pi$ system is mainly produced with orbital angular momentum $l=1$,
kinematically as if the two-pions were arising from the low mass
tail of the $\rho$-meson! To make this hypothesis more quantitative, 
consider the simplest matrix element for the production of a $p$-wave
$\pi\pi$ pair, which is
\begin{equation}
\label{me}
M=\sqrt{3}C\,\bar{u}_{\tau}\,\vec{\epsilon}\cdot(\hat{K}\times\vec{k})\,u_{p}
\:,
\end{equation}
where $\vec{\epsilon}$ is the deuteron polarisation vector,
$u_{p}$ and $u_{\tau}$ are Pauli spinors describing respectively the 
initial proton and final $^3$He, and $C$ is a constant.
The beam momentum is denoted by $\vec{K}$ and the relative momentum of
the two pions as $\vec{k}=\frac{1}{2}(\vec{k}_1-\vec{k}_2)$.

This matrix element only allows pion pairs with $m_{l}=\pm 1$ along the
beam direction and, squaring and averaging over spins, leads to
\begin{equation}
\overline{|M|^2} = |C|^2\,|\hat{K}\times\vec{k}|^2 =
k^2|C|^2\sin^2\theta_{\pi\pi p}\:,
\end{equation}
which immediately reproduces the angular dependence observed in fig.~2d.

After averaging over $\theta_{\pi\pi p}$, the differential cross
section becomes
\begin{equation}
\label{cross}
d\sigma = \overline{|M|^2}\:dLips = \frac{2}{3}|C|^2k^2\:dLips\:,
\end{equation}
where {\it dLips} is the Lorentz-invariant phase space. Since
non-relativistically $T_{\pi\pi}= k^2/m_{\pi}$, eq.~(\ref{cross})
immediately leads to the observed proportionality in $T_{\pi\pi}$ times 
phase space. This is shown as the solid curve in fig.~2a, which
however was calculated including the small relativistic effects.

Despite the $\pi\pi$ $p$-wave hypothesis resulting in a large deviation
from phase space for the distribution in pion-pion excitation
energies, the modifications to the $\pi$-$^3$He relative energy are
extremely small and would actually vanish in the limit of infinite 
$^3$He mass. This agrees well with the results shown in fig.~2b. Note
that in our data cannot distinguish between the energy distributions of
the $\pi^+$ and $\pi^-$.

Experimentally the pions are preferentially emitted at large angles
with respect to the beam direction and the $p$-wave hypothesis leads
to an effect of this kind, though not quite as large as that implied
by the experimental data shown in fig.~2c.

Although $s$-wave two-pion production has been observed in the
$pd\to\,^{3}\mbox{\rm He}\,\pi^+\pi^-$ reaction
very close to threshold~\cite{Vigdor}, and at $Q\approx 200$~MeV $s$-wave
pion-pion pairs again dominate the spectrum through the ABC enhancement,
it is clear from the present data that there is intermediate $Q$-range where
$p$-waves are dominant. Similar behaviour is, however, observed in other
experiments. The missing-mass distributions obtained for the $np\to d X$
reaction at $Q\approx 200$~MeV show a striking ABC effect~\cite{Plouin},
whereas at 70~MeV no ABC is seen~\cite{Hollas}. In the latter case the
events are pushed to the maximum missing mass, which is consistent with
the $p$-wave production seen in fig.~2a. Furthermore, recent data on the
comparison of pion production in the $\pi^+d\to \pi^+\pi^+nn$ and  
$\pi^+d\to \pi^+\pi^-pp$ reactions at $Q\approx 100$~MeV show that,
whereas the $\pi^+\pi^+$ spectrum broadly follows phase space
modulated by detector acceptance, the $\pi^+\pi^-$ data are again
heavily biased towards the maximum value of $T_{\pi\pi}$~\cite{Chaos}.

Kinematically our results are indistinguishable from the
production of the low-mass part of the $\rho$-meson in 
$pd\to\,^{3}\mbox{\rm He}\,\rho^0$, with the $\rho$-mesons being formed
with polarizations $\pm 1$ in the beam direction. Though there is some
evidence from photoproduction for the $\rho$ mass being depressed in the
mass-3 system~\cite{TAGX}, it is hard in our case to see why such 
production should become less important at the higher energies
where the original inclusive measurements were performed~\cite{ABC}. One
possibility is that the effect is due to a rare decay of the $\Delta$ isobar
since at $Q=70$~MeV the invariant mass with respect to a single nucleon
is only 1290~MeV, which is well within the $\Delta$ width. Though favoured
by isospin, and also offering a natural explanation for the non-observance
at high $Q$, any dynamical model based on this idea would have to transfer
one unit of angular momentum from the final to the initial state through
the action of a recoil term.

Conventional models of ABC production~\cite{Risser,Franz,GRD} suggest 
that this arises through two independent $p$-wave pion productions, 
mediated by two $\Delta$ resonances, combining to give $s$-wave pion-pion
pairs. At low energies one of these productions might be through an
$s$-wave $\pi N$ system, leaving only one unit of angular momentum in the
final state. However, given the importance of the $\Delta$ almost down to
threshold, this is unlikely to play a major role here and, in any
case, would tend to lead to $p$-waves between the pion and the $^3$He. 

Further clarification of the reaction mechanism will be provided
through a study of the variation of the 
$pd\to\,^{3}\mbox{\rm He}\,\pi^+\pi^-$ cross sections with beam
energy, data on which are currently being analyzed. Since all events
with a $^{3}\mbox{\rm He}$ in the focal plane have been written on tape,
it is furthermore hoped to deduce the $\pi^0\pi^0$ excitation spectra
through a subtraction of the exclusive data from that obtained inclusively
purely using the spectrometer.

Experiments are currently being performed on exclusive
$pd\to\,^{3}\mbox{\rm He}\,K^+K^-$ production at similar $Q$-values
and it will be very interesting to see whether the $\phi$'s produced
in this reaction have a similar alignment to that observed for the
$p$-wave pion-pion pairs.

We wish to thank the COSY crew for providing the high quality beam and
the Big Karl technical staff for their untiring efforts. This work was
supported by the Bundesministerium f\"ur Bildung und Wissenschaft and
the IKP J\"ulich.


\begin{figure}
\epsfig{figure=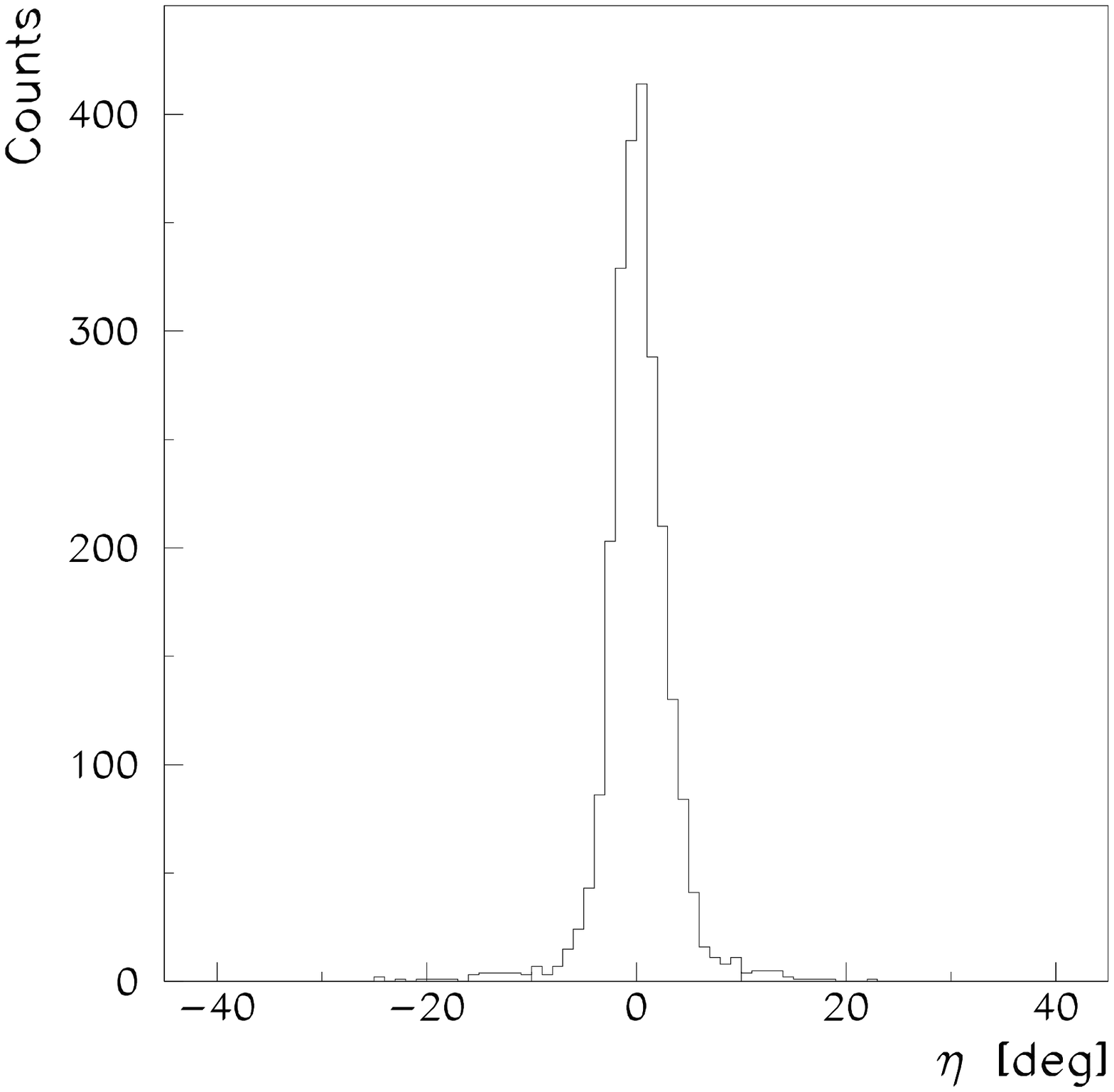, width=14cm}
\caption{Distribution in coplanarity angle $\eta$. Here 
$\eta$ is the deviation of the missing momentum, defined by Big Karl,
and the plane, defined by the directions of the two hits
in the MOMO detector.}
\end{figure}
\newpage
\begin{figure}
\epsfig{figure=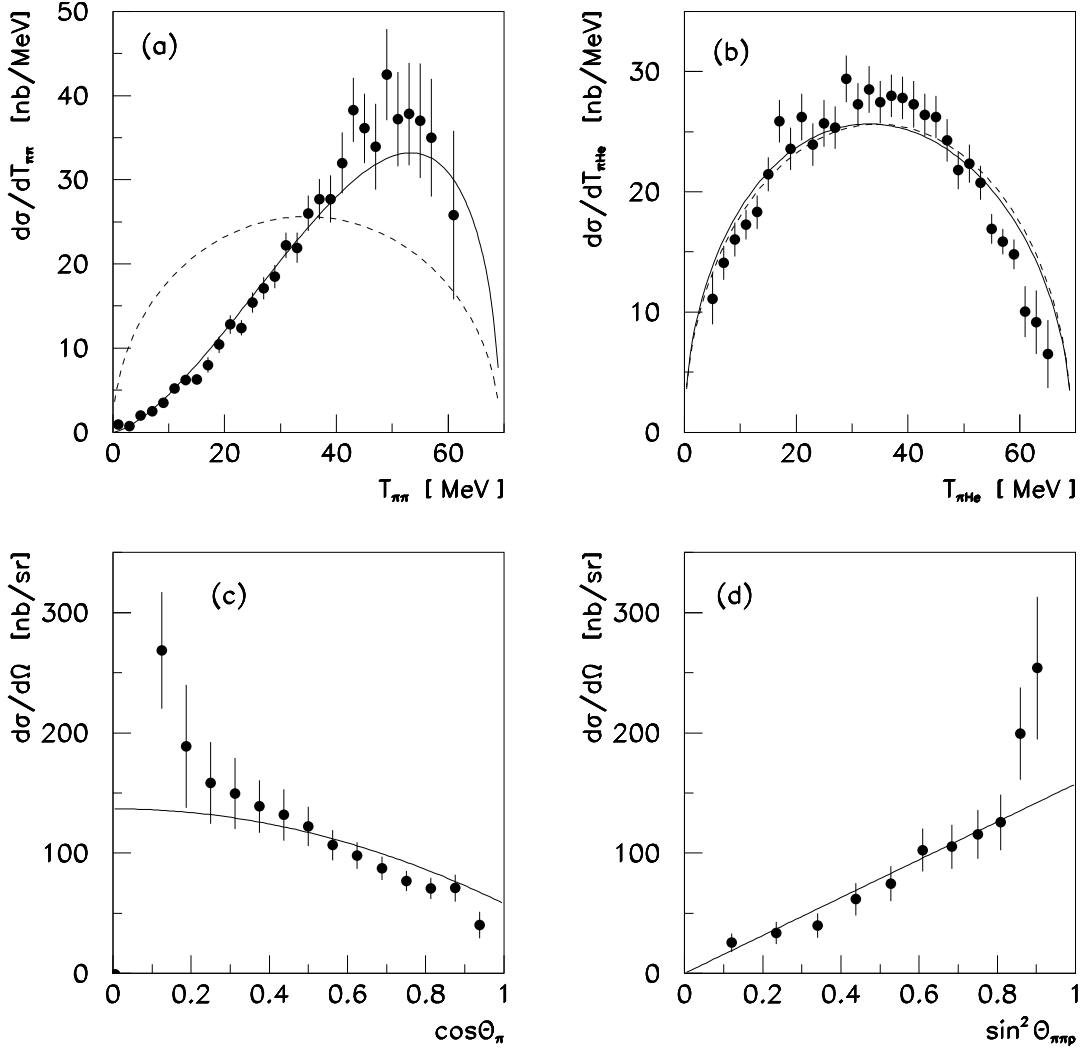, width =16cm}
\caption{Differential cross sections for the 
$pd\to\, ^{3}\mbox{\rm He}\,\pi^+\pi^-$ reaction at $T_p=546$~MeV as a
function of (a) the pion-pion excitation energy $T_{\pi\pi}$, (b) the
excitation energy in the $\pi\,^3$He system, (c) the
angle $\theta_{\pi}$ between one of the pions and the beam direction
in the overall c.m.\ system, and (d) the angle $\theta_{\pi\pi p}$ 
between the two-pion relative momentum and the beam direction, also in
the c.m.\ system. In the first two cases the dashed curves represent the
predictions of phase space normalised to the data, whereas in all
cases the solid curves are predictions assuming that the pion-pion pair
emerges in the relative $p$-wave described by the matrix element of 
eq.~(\ref{me}). The linear deviations in  $T_{\pi\pi}$ from phase
space  in (a) and the linearity of the cross section with 
$\sin^2\theta_{\pi\pi p}$ in (d) are clear indications of pion-pion $p$-wave
effects.}
\end{figure}

\end{document}